\documentclass[sigconf, nonacm]{acmart}
\usepackage{xspace}
\usepackage{booktabs}   
\usepackage{tabularx}   
\usepackage{graphicx}   

\graphicspath{{images/}} 

\newcommand{\MLMVS}{Memory-locked synthesis\xspace}
\newcommand{\ECGD}{Evidence-coverage--guided execution\xspace}
\newcommand{\SPLCD}{Section-packed long-context grounding\xspace}

\AtBeginDocument{%
  }

\setcopyright{acmlicensed}
\copyrightyear{2026}
\acmYear{2026}
\acmDOI{XXXXXXX.XXXXXXX}

\acmConference[SIGIR '26]{Proceedings of the 49th International ACM SIGIR Conference on Research and Development in Information Retrieval}{July 21--25, 2026}{TBD}
\acmPrice{15.00}
\acmISBN{978-1-4503-XXXX-X/26/07}

\begin{document}

\title{Orchestrating Specialized Agents for Trustworthy Enterprise RAG}

\author{Xincheng You}
\email{xyou@atlassian.com}
\affiliation{%
  \institution{Atlassian}
  \city{Cambridge}
  \country{USA}
}
\author{Qi Sun}
\email{qsun@atlassian.com}
\affiliation{%
  \institution{Atlassian}
  \city{Mountain View}
  \country{USA}
}
\author{Neha Bora}
\email{nbora@atlassian.com}
\affiliation{%
  \institution{Atlassian}
  \city{Washington DC}
  \country{USA}
}
\author{Huayi Li}
\email{hli8@atlassian.com}
\affiliation{%
  \institution{Atlassian}
  \city{Mountain View}
  \country{USA}
}
\author{Shubham Goel}
\email{sgoel8@atlassian.com}
\affiliation{%
  \institution{Atlassian}
  \city{San Francisco}
  \country{USA}
}
\author{Kang Li}
\email{kli3@atlassian.com}
\affiliation{%
  \institution{Atlassian}
  \city{Bellevue}
  \country{USA}
}

\author{Sean Culatana}
\email{sculatana@atlassian.com}
\affiliation{%
  \institution{Atlassian}
  \city{Mountain View}
  \country{USA}
}

\renewcommand{\shortauthors}{You et al.}

\begin{abstract}
Retrieval-Augmented Generation (RAG) shows promise for enterprise knowledge work, yet it often underperforms in high-stakes decision settings that require deep synthesis, strict traceability, and recovery from underspecified prompts. One-pass retrieval-and-write pipelines frequently yield shallow summaries, inconsistent grounding, and weak mechanisms for completeness verification.

\noindent We introduce ADORE (Adaptive Deep Orchestration for Research in Enterprise), an agentic framework that replaces linear retrieval with iterative, user-steered investigation coordinated by a central orchestrator and a set of specialized agents. ADORE’s key insight is that a structured Memory Bank (a curated evidence store with explicit claim--evidence linkage and section-level admissible evidence) enables traceable report generation and systematic checks for evidence completeness.

\noindent Our contributions are threefold: (1) \emph{\MLMVS}—report generation is constrained to a structured Memory Bank (Claim--Evidence Graph) with section-level admissible evidence, enabling traceable claims and grounded citations; (2) \emph{\ECGD}—a retrieval--reflection loop audits section-level evidence coverage to trigger targeted follow-up retrieval and terminates via an evidence-driven stopping criterion; (3) \emph{\SPLCD}—section-level packing, pruning, and citation-preserving compression make long-form synthesis feasible under context limits.

\noindent Across our evaluation suite, ADORE ranks first on DeepResearch Bench ($52.65$) and achieves the highest head-to-head preference win rate on DeepConsult ($77.2\%$) against commercial systems.
\end{abstract}



\maketitle

\section{Introduction}
Large Language Models (LLMs) are increasingly embedded in enterprise knowledge workflows, reshaping how employees discover, summarize, and operationalize institutional information. Retrieval-Augmented Generation (RAG) \cite{lewis2020retrieval} is the dominant grounding pattern: it ties generation to a retrieved evidence set, aiming to reduce hallucination and improve factuality on proprietary corpora \cite{gao2023retrieval, ji2023survey}. In practice, however, enterprise queries are rarely clean ``one-shot'' QA. They are underspecified, multi-faceted, and often require iterative decomposition, cross-document synthesis, and auditable traceability.

\noindent Despite strong performance on short-form QA, many ``retrieve-then-generate'' pipelines fall short for executive decision support. First, they can confidently answer the wrong question when the prompt is ambiguous, producing plausible but misaligned narratives \cite{gao2023retrieval, ji2023survey}. Second, linear pipelines are brittle: when early retrieval misses a critical subtopic, downstream generation often compounds the omission. Third, long contexts are not a panacea; even when relevant evidence is present, models can ignore or underuse it (the ``lost in the middle'' effect) \cite{liu2024lost}. These issues become acute for long-form reporting, where completeness, citation stability, and section coverage matter as much as fluency.

\paragraph{Limitations of static RAG in enterprise.}
Our analysis of real-world enterprise deployment reveals three recurring failure modes:
\begin{itemize}
    \item \textbf{Prompt ambiguity \& cold start.} Users seldom phrase precise, decontextualized queries. Requests like ``Analyze our Q3 risks'' are informationally underspecified. Without clarification behaviors \cite{aliannejadi2019asking}, systems retrieve against vague lexical cues, reducing precision and recall.
    \item \textbf{Workflow rigidity.} Treating retrieval as a single pre-processing step limits recovery from early mistakes. Static ``chains'' often lack the agency to adjust strategy when intermediate evidence indicates a mismatch or missing dimension \cite{yao2023react}.
    \item \textbf{Contextual myopia.} Long-context ingestion does not guarantee utilization; important details can be ignored even when retrieved \cite{liu2024lost}. Moreover, one-pass synthesis lacks self-correction loops that can systematically target missing or weakly supported sections \cite{madaan2024self, asai2024selfrag}.
\end{itemize}

\paragraph{Positioning vs.\ prior deep-research agents.}
ADORE is inspired by recent deep-research systems, but it differs in what drives iteration and what constitutes convergence.
WebWeaver emphasizes outline-centric evidence structuring and section-level writing with a memory bank \cite{li2025webweaver}.
TTD-DR frames long-form generation as test-time diffusion over drafts, improving quality via iterative denoising steps \cite{han2025ttddr}.
In contrast, ADORE introduces \emph{\MLMVS}: we treat the Memory Bank as a hard constraint on generation (section-scoped admissible evidence with explicit claim--evidence linkage), then use section-level evidence coverage audits to diagnose gaps and drive targeted follow-up retrieval with an evidence-driven stopping rule.
This mechanism is operationalized via \ECGD, with an explicit evidence-coverage stopping rule rather than a fixed iteration budget or outline-only convergence.

\paragraph{ADORE overview.}
ADORE (Adaptive Deep Orchestration for Research in Enterprise) replaces linear ``retrieve-then-write'' with an agentic workflow that mirrors human research: scoping via clarification, plan formation, iterative evidence collection, and verification. The core design choice is a structured Memory Bank that stores evidence with explicit claim--evidence linkage, enabling traceable report generation and systematic checks for coverage and citation stability.

\begin{figure}[t]
  \centering
  \includegraphics[width=\columnwidth]{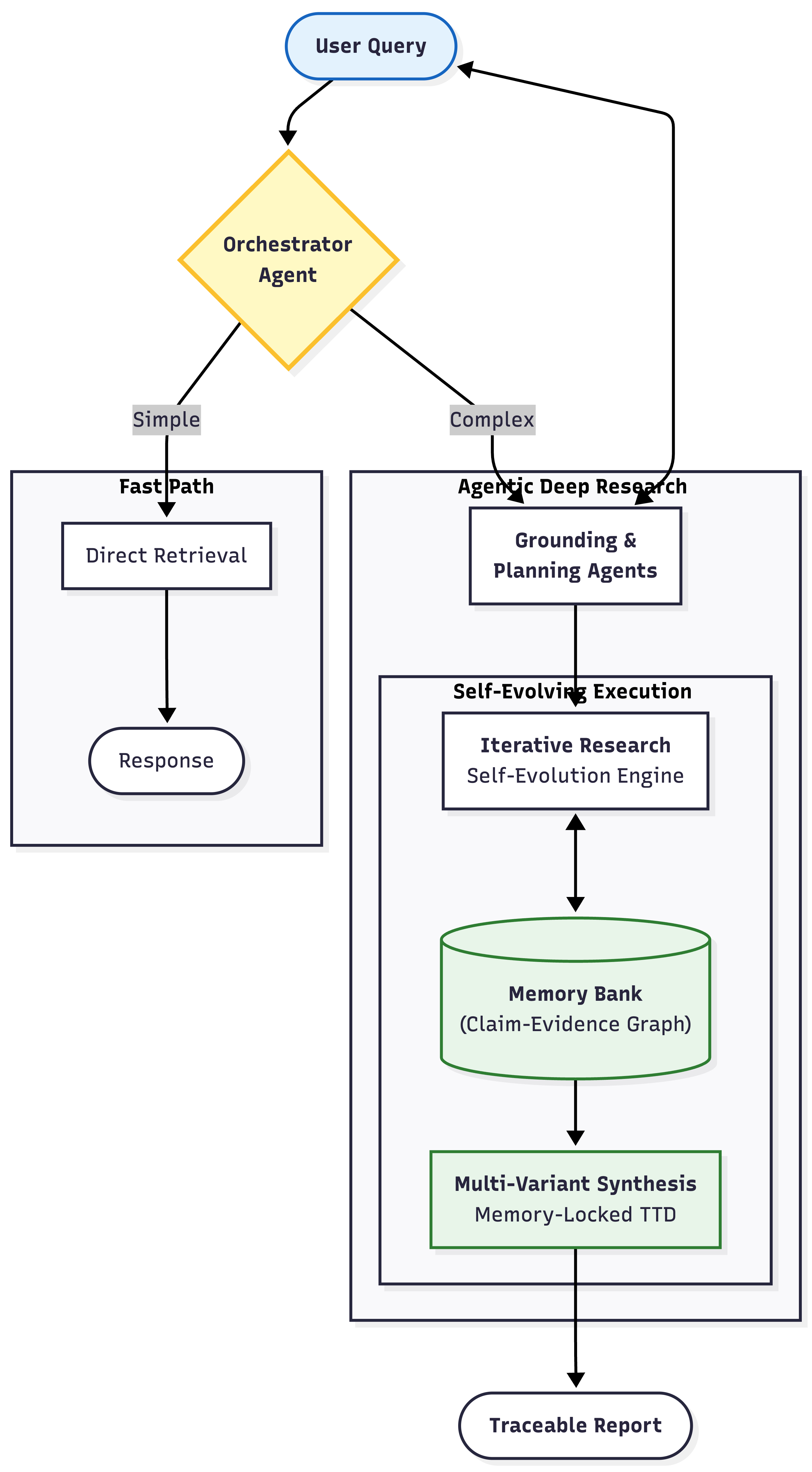}
  \caption{The ADORE framework. The Orchestrator routes complex tasks to an agentic workflow that replaces linear retrieval with \MLMVS. Unlike standard RAG, ADORE constrains generation to a structured Memory Bank (Claim--Evidence Graph), enabling iterative, evidence-coverage--guided execution for traceable enterprise reporting.}
  \Description{High-level ADORE architecture: an orchestrator routes to either a fast retrieval path or a deep research workflow with grounding, planning, iterative execution, a memory bank, and traceable report output.}
  \label{fig:teaser}
\end{figure}

\paragraph{Contributions.}
We make the following contributions (using consistent terminology throughout the paper):
\begin{enumerate}
    \item \textbf{\MLMVS.}~We generate long-form reports \emph{constrained to a structured Memory Bank} (Claim--Evidence Graph with section-level admissible evidence), ensuring that claims are traceable and citations are grounded in retrieved sources.
    \item \textbf{\ECGD.}~We use section-level evidence coverage audits to localize weak or missing sections, trigger targeted follow-up retrieval, and terminate via an evidence-driven stopping criterion \cite{asai2024selfrag, trivedi2023ircot}.
    \item \textbf{\SPLCD.}~We make long-form grounded reporting tractable by packing section-specific evidence, pruning redundant context, and compressing retrieved content into citation-preserving summaries so synthesis stays within context limits \cite{liu2024lost, beltagy2020longformer}.
    \item We implement an adaptive orchestrator that routes between fast retrieval and deep multi-agent investigation, and we demonstrate state-of-the-art results on both public and internal benchmarks \cite{du2025deepresearch, deepconsult2025}.
\end{enumerate}

\section{The ADORE Framework}

ADORE abandons a single linear pipeline in favor of a hub-and-spoke architecture governed by a central \textbf{Orchestrator Agent} (Figure~\ref{fig:teaser}). The orchestrator continuously interprets user intent, conversation context, and intermediate evidence signals to select an execution path: a low-latency retrieval response for simple informational queries, or a multi-agent deep-research workflow for complex analytical tasks.

\subsection{Adaptive Orchestration and Routing}
The Orchestrator classifies incoming requests by \emph{research complexity} (e.g., ambiguity, expected depth, number of sub-questions, and required traceability). Simple factoid-style requests follow a standard retrieval path. Complex requests trigger a multi-phase workflow:
\begin{itemize}
    \item \textbf{Clarification (Grounding):} resolve ambiguity and establish success criteria before heavy retrieval starts.
    \item \textbf{Planning:} produce a structured research plan that the user can accept or edit.
    \item \textbf{Execution:} run iterative retrieval, synthesis, and verification until evidence coverage criteria are satisfied.
\end{itemize}

\noindent A key principle is that \emph{iteration is evidence-driven}. Rather than iterating a fixed number of times, ADORE uses \ECGD\ signals (coverage and stability) to decide whether to retrieve more, refine the outline, or finalize.

\subsection{Specialized Agents}
ADORE delegates responsibilities to specialized agents:
\begin{itemize}
    \item \textbf{Grounding Agent:} mitigates cold start by formulating clarification questions (and proposing initial scope assumptions) to convert underspecified prompts into a concrete brief \cite{aliannejadi2019asking}.
    \item \textbf{Planning Agent:} converts the refined brief into a structured plan (sections, priorities, success criteria). The plan is presented as a shared artifact so users can edit or prioritize sections, keeping humans in the loop.
    \item \textbf{Execution Agent:} manages retrieval and synthesis. It uses a \textbf{Self-Evolution Engine} to evolve search queries based on intermediate findings and detected evidence gaps, enabling multi-hop recovery \cite{trivedi2023ircot}.
    \item \textbf{Report Generation Agent:} produces (i) an initial outline, (ii) outline revisions based on retrieved evidence, and (iii) final reports with traceable citations.
    \item \textbf{WebSearch Agent:} retrieves and filters high-authority public sources when external context is required, discarding low-signal content before ingestion.
\end{itemize}

\subsection{Execution Workflow}
The \textbf{Execution Agent} orchestrates the research loop as follows:
\begin{enumerate}
    \item \textbf{Outline draft:} The Report Generation Agent proposes a report outline aligned with the user-confirmed plan.
    \item \textbf{Iterative retrieval:} The Self-Evolution Engine proposes search queries over internal and external sources. A reflection step audits retrieved evidence against the outline to identify missing sections and weak support; queries are refined and re-executed.
    \item \textbf{Outline update:} The Report Generation Agent revises the outline using newly retrieved evidence.
    \item \textbf{Evidence-driven stopping:} The Execution Agent checks whether evidence coverage and citation stability satisfy the plan. If not, it returns to retrieval; otherwise, it triggers final report generation.
\end{enumerate}

\begin{figure}[t]
  \centering
  \includegraphics[width=\columnwidth]{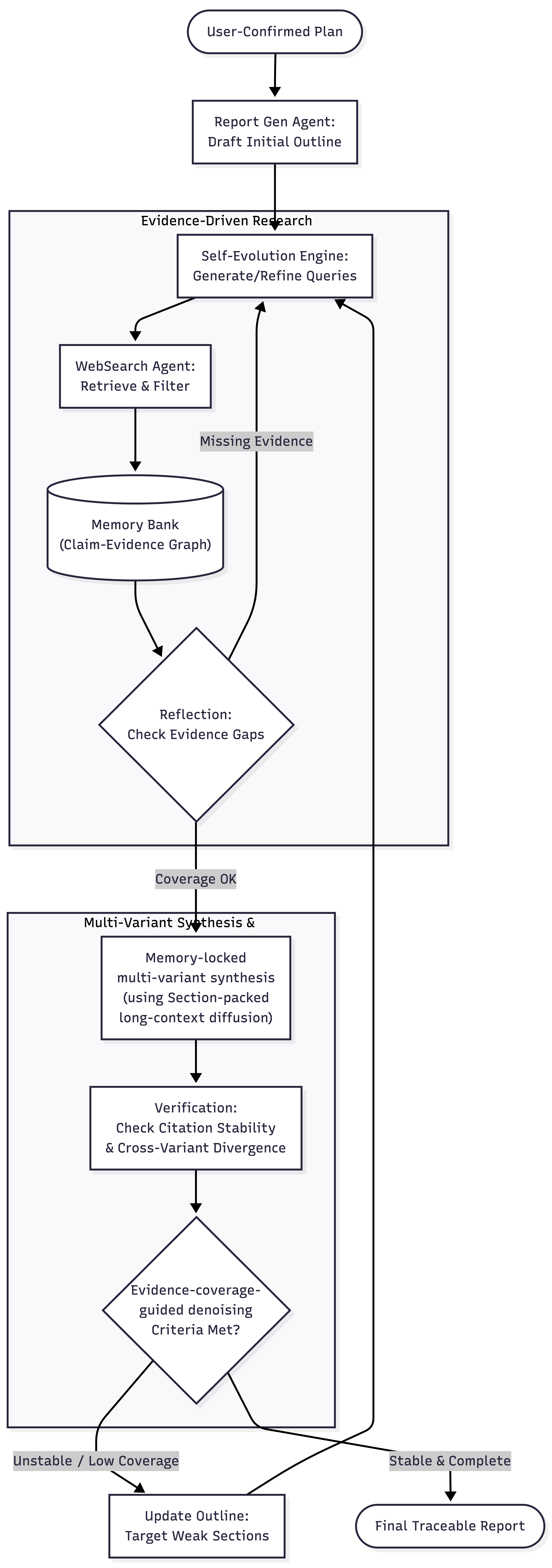}
  \caption{Self-evolving execution with \MLMVS. The workflow iterates through retrieval and reflection, storing evidence in a structured Memory Bank. The process uses section-level evidence coverage audits to localize weak or missing sections, triggering targeted follow-up retrieval and terminating when coverage criteria are met.}
  \Description{Flowchart of the execution loop: outline drafting, iterative retrieval and reflection, evidence stored in a Memory Bank, coverage audits triggering targeted retrieval, and final traceable report generation.}
  \label{fig:execution}
\end{figure}

\subsection{Synthesis and Traceability}

ADORE draws inspiration from diffusion-style iterative drafting for long-form reports \cite{han2025ttddr} and outline-grounded writing with structured evidence memories \cite{li2025webweaver}. Our departure is to make the Memory Bank a \emph{hard constraint} on generation and to use explicit section-level evidence coverage signals to drive targeted retrieval and determine convergence.

\subsubsection{Memory Bank (Claim--Evidence Graph)}
The Memory Bank is a persistent evidence store with explicit claim--evidence linkage. For each planned section, it maintains the admissible evidence set (sources and excerpts) and tracks coverage requirements (what must be supported). This structure ensures that report writing is grounded and that any claim can be traced to supporting documents.

\subsubsection{\MLMVS}
Given a fixed Memory Bank, ADORE generates a report under a strict admissible-evidence constraint: each section is written using only the section-scoped evidence stored in the claim--evidence graph. This enables traceability by construction (each claim must map to supporting citations in the Memory Bank) and supports citation auditing by verifying that cited sources and quoted spans appear in the admissible evidence store.

\subsubsection{\ECGD\ Criterion}
ADORE drives iteration via section-level evidence coverage audits. After each retrieval--reflection cycle, the system evaluates whether each planned section is sufficiently supported by admissible evidence in the Memory Bank. When coverage indicates insufficient support, the system triggers targeted follow-up retrieval and focused rewriting for the affected sections. This yields an \emph{evidence-driven stopping criterion}: the loop terminates when coverage requirements satisfy the plan, rather than after a fixed number of refinement steps \cite{asai2024selfrag, trivedi2023ircot}.

\subsubsection{\SPLCD\ Strategy}
Long-form grounded reporting is context-intensive. \SPLCD\ makes it tractable by:
(i) packing only section-relevant evidence into each generation call,
(ii) pruning redundant or low-salience retrieved text, and
(iii) compressing long sources into citation-preserving summaries so synthesis remains grounded without exceeding context budgets \cite{liu2024lost, beltagy2020longformer}.
This preserves traceability while keeping long-form generation practical at enterprise scale.

\section{Evaluation Methodology}

We employ a dual evaluation strategy combining reference-based metrics and reference-free preference judgments.

\subsection{Reference-Based: The RACE Framework}
We utilize the \textbf{RACE} framework (Readability, Instruction Following, Comprehensiveness, Insight) to benchmark report quality \cite{du2025deepresearch}. RACE generates a prompt-specific weight vector $W_T$ tailored to task requirements.

\noindent For a given research task $T$, RACE generates a weight vector $W_T$ tailored to the specific requirements of the prompt. For example, a task analyzing ``Provide a comprehensive market analysis of Strategic Portfolio Management (SPM) technology solutions'' assigns higher weight to Comprehensiveness:
\begin{equation}
W_{Insight} = 0.3, \quad W_{Comp} = 0.35, \quad W_{Instruct} = 0.2, \quad W_{Read} = 0.15
\end{equation}

\noindent Within the \textbf{comprehensiveness} dimension, RACE generated detailed sub-criteria with differentiated weights. For instance, for the SPM market-analysis task the generated rubrics and their assigned weights were:
\begin{table}[t]
\centering
\small
\caption{Comprehensiveness sub-criteria and weights for the SPM market-analysis task.}
\label{tab:race_comp_weights}
\begin{tabularx}{\columnwidth}{@{}Xr@{}}
\toprule
\textbf{Sub-criterion} & \textbf{Weight} \\
\midrule
Definition and Conceptual Framework of SPM & 10\% \\
Market Landscape Analysis & 15\% \\
Technical Features and Capabilities Coverage & 15\% \\
Vendor Assessment Comprehensiveness & 15\% \\
Selection Decision Factors Analysis & 10\% \\
Evaluation Framework Robustness & 15\% \\
Implementation and Best Practices Coverage & 15\% \\
\bottomrule
\end{tabularx}
\end{table}

\subsection{Reference-free: Side-by-Side Evaluation}
In addition to reference-based benchmarks, we run side-by-side evaluations against competitor deep research products. For each research task, we:
\begin{enumerate}
 \item Generate a report using ADORE.
 \item Obtain a competing report (e.g., ChatGPT Deep Research).
 \item Compare the two reports using an LLM-judge win/tie/lose framework.
\end{enumerate}
This setup estimates how often users or judges would prefer ADORE when viewing reports side by side.

\begin{table*}[t]
  \centering
  \small
    \caption{RACE scores on DeepResearch Bench: ADORE compared with leading deep-research systems.}
  \label{tab:deepresearch_bench}
  \begin{tabular}{llccccc}
    \toprule
    \textbf{Rank} & \textbf{Model} & \textbf{Overall} & \textbf{Comp.} & \textbf{Insight} & \textbf{Inst.} & \textbf{Read.} \\
    \midrule
    \textbf{1} & \textbf{ADORE (Ours)} & \textbf{52.65} & \textbf{52.22} & \textbf{54.37} & \textbf{51.11} & \textbf{52.18} \\
    2 & Tavily Research & 52.44 & 52.84 & 53.59 & 51.92 & 49.21 \\
    3 & ThinkDepthAI & 52.43 & 52.02 & 53.88 & 52.04 & 50.12 \\
    4 & CellCog & 51.94 & 52.17 & 51.90 & 51.37 & 51.94 \\
    5 & Salesforce Air & 50.65 & 50.00 & 51.09 & 50.77 & 50.32 \\
    6 & LangChain (GPT-5) & 50.60 & 50.06 & 50.76 & 51.31 & 49.72 \\
    7 & Gemini 2.5 Pro & 49.71 & 49.51 & 49.45 & 50.12 & 50.00 \\
    8 & LangChain (Tavily) & 49.33 & 49.80 & 47.34 & 51.05 & 48.99 \\
    9 & OpenAI Deep Research & 46.45 & 46.46 & 43.73 & 49.39 & 47.22 \\
    10 & Claude Research & 45.00 & 45.34 & 42.79 & 47.58 & 44.66 \\
    \bottomrule
  \end{tabular}
\end{table*}

\begin{table*}[ht!]
  \centering
  \small
    \caption{DeepConsult side-by-side evaluation: win/tie/lose rates and aggregate score for ADORE vs.\ agentic baselines.}
  \label{tab:eval_results}
  \begin{tabular}{llcccc}
    \toprule
    \textbf{Rank} & \textbf{Agentic System} & \textbf{Win (\%)} & \textbf{Tie (\%)} & \textbf{Lose (\%)} & \textbf{Score} \\
    \midrule
    \textbf{1} & \textbf{ADORE (Ours)} & \textbf{77.21} & \textbf{18.38} & \textbf{4.41} & \textbf{7.03} \\
    2 & Enterprise Deep Research & 71.57 & 19.12 & 9.31 & 6.82 \\
    3 & WebWeaver (Claude-sonnet) & 66.86 & 10.47 & 22.67 & 6.96 \\
    4 & WebWeaver (gpt-oss-120b) & 65.31 & 11.22 & 23.47 & 6.64 \\
    5 & Gemini-2.5-pro & 61.27 & 31.13 & 7.60 & 6.70 \\
    6 & WebWeaver (qwen3-235b) & 54.74 & 28.61 & 16.67 & 6.47 \\
    7 & openai-deepresearch & 0.00 & 100.00 & 0.00 & 5.00 \\
    8 & Perplexity Deep Research & 32.00 & - & - & - \\
    9 & doubao-research & 29.95 & 40.35 & 29.70 & 5.42 \\
    10 & Claude-research & 25.00 & 38.89 & 36.11 & 4.60 \\
    11 & WebWeaver (qwen3-30b) & 28.65 & 34.90 & 36.46 & 4.57 \\
    12 & WebShaper (32B) & 3.25 & 3.75 & 93.00 & 1.63 \\
    \bottomrule
  \end{tabular}
\end{table*}

\subsection{Evaluation Datasets}
We evaluate ADORE on three datasets:
\begin{enumerate}
    \item \textbf{DeepResearch Bench (Public):} 100 PhD-level research tasks authored by domain experts (reference-based) \cite{du2025deepresearch}.
    \item \textbf{Internal Enterprise Benchmark:} proprietary tasks derived from high-quality internal documents via reverse prompt engineering (reference-based).
    \item \textbf{Deep Consult Dataset (Public):} business/consulting research tasks (reference-free) \cite{deepconsult2025}.
\end{enumerate}

\section{Experimental Results}

\subsection{RACE Performance}
On the public \textbf{DeepResearch Bench}, ADORE achieved a score of \textbf{52.65}, with the following component scores:
\begin{itemize}
    \item Comprehensiveness: \textbf{52.22}
    \item Insight: \textbf{54.37}
    \item Instruction Following: \textbf{51.11}
    \item Readability: \textbf{52.18}
\end{itemize}
This places it at rank \#1 on the leaderboard \cite{du2025deepresearch}. Table~\ref{tab:deepresearch_bench} compares our performance against top industry baselines.

\noindent On the \textbf{Internal Enterprise Benchmark}, the system achieved a RACE score of \textbf{64.11}, with the following component scores:
\begin{itemize}
    \item Comprehensiveness: \textbf{65.91}
    \item Insight: \textbf{66.76}
    \item Instruction Following: \textbf{60.35}
    \item Readability: \textbf{61.95}
\end{itemize}
Given that the benchmark is calibrated such that a score of 50.0 represents the human-written reference, ADORE demonstrates a \textbf{14.11\%} improvement in quality over human baselines for these specific enterprise tasks.

\subsection{Side-by-Side Preference}
We conducted blind side-by-side evaluation on \textbf{DeepConsult} \cite{deepconsult2025}. We compared ADORE against a leading competitor (ChatGPT Deep Research) using an LLM-as-a-Judge paradigm.

\noindent As shown in Table~\ref{tab:eval_results}, ADORE achieves a win rate of \textbf{77.21\%} with a lose rate of \textbf{4.41\%}.

\section{Conclusion}

We presented ADORE, an agentic framework for deep enterprise research that moves beyond rigid RAG pipelines toward adaptive, evidence-driven investigation.
ADORE’s central novelty is operational: \MLMVS\ constrains report generation to a structured Memory Bank for traceability; \ECGD\ turns evidence coverage into targeted retrieval decisions with a stopping rule; and \SPLCD\ keeps long-form grounded synthesis practical under long-context constraints.
Across public and internal benchmarks, ADORE achieves state-of-the-art results while maintaining traceable, section-level evidence grounding.

\noindent Future work will focus on:
\begin{itemize}
    \item \textbf{Actionability:} extending planning to propose decision options and trade-offs, not only synthesis.
    \item \textbf{Factual guardrails:} developing granular citation-accuracy metrics and automatic audits for technical domains.
    \item \textbf{Multimodal synthesis:} generating visual artifacts (charts, plots) aligned with evidence-backed report sections.
\end{itemize}

\begin{acks}
We thank the GenAI platform, AI fundamental, assistant evaluation teams for their support on the project.
\end{acks}

\bibliographystyle{ACM-Reference-Format}
\bibliography{sample-base}

\end{document}